# Room-temperature tunnel current amplifier and experimental setup for high resolution electronic spectroscopy in millikelvin STM experiments

Hélène le Sueur and Philippe Joyez
Service de Physique de l'Etat Condensé (CNRS URA 2464)
CEA Saclay
91191 Gif sur Yvette Cedex, France

**abstract**
The spectroscopic resolution of tunneling measurements performed with a scanning tunneling microscope is ultimately limited by the temperature at which the experiment is performed. To take advantage of the potential high spectroscopic resolution associated with operating an STM in a dilution refrigerator we have designed a room temperature tunnel current amplifier having very small back-action on the tunnel contact and allowing to nearly reach the predicted energy resolution. This design is a modification of the standard op-amp based tip-biasing current-voltage converter which implements differential voltage sensing and whose back action on the tip voltage is only ~2 μV rms for a 14 MV/A transimpedance and 22kHz bandwidth.

**Introduction**
Scanning tunneling microscopes (STM) operated in spectroscopy mode provides valuable local information on the energy dependence of electronic mechanisms in the sample. For this purpose, low temperature STMs are in particular widely used to investigate recently discovered superconductors, or proximity effects. In this tunneling spectroscopy technique, when one of the electrodes is non-superconducting, the energy resolution is ultimately limited by the temperature at which the experiment is carried out. Indeed, the most precise results in this field are obtained using STMs in dilution refrigerators [1-7] with which one can commonly cool experiments in the 10-100 mK range. However, in these experiments the energy resolution usually achieved is in practice significantly less than the thermal fundamental limit. There is thus a need for better instrumentation allowing to fully exploit the potential of these low temperature measurements.

In this article we discuss the required characteristics of the experimental setup in order to achieve optimal spectroscopic resolution and we present a room temperature amplifier optimised to have a back-action noise on the tunnel contact low enough that the energy resolution is close to the temperature limited value in an experiment carried out at 40 mK.

**Tunnel spectroscopy basics**
For a perfect voltage bias between two electrodes made of good metals, the tunnel current can be expressed as

$$I(V) = \frac{4\pi e |\tau|^2}{\hbar} \int (n_L(\epsilon - eV) - n_R(\epsilon)) \rho_L(\epsilon - eV) \rho_R(\epsilon) d\epsilon$$

where $n_{L,R}(\epsilon)$ are the occupancies of the electronic states at energy ε in the left or right electrode, $\rho_{L,R}(\epsilon)$ are the densities of states in the electrodes and τ is the tunneling Hamiltonian matrix element between the two electrodes, assumed here to be independent of energy. When one knows the density of states and the occupation of the energy levels in one of the electrodes, the measurement of the *I-V* characteristic of the contacts allows to extract information on the electronic state of the second electrode. For instance, if the left electrode is a normal metal in equilibrium at zero temperature, $\rho_L$ can be assumed nearly constant near the fermi level and $n_L$ is a step function at the fermi energy. Then, the derivative $dI(V)/dV$, gives a direct measurement of $\rho_R(eV)$. If the temperature $T$ is finite, the step function becomes a Fermi function, *i.e.* a rounded step function. In this case, the differential conductance $dI(V)/dV$ is given by the convolution product $\rho_R * f'_L$ :

$$\frac{dI(V)}{dV} \propto \int_{-\infty}^{\infty} \rho_R(\epsilon) f'_L(eV-\epsilon) d\epsilon \qquad (1)$$

where $f'_L$ is the derivative of the Fermi function $f_L$ of the left electrode which is a bell-shaped function having a full width at half maximum of $2\ln(3+2\sqrt{2})kT$. Such a convolution product smears out the details in $\rho_R$, limiting the energy resolution to $\sim 3\ kT$.

In real life experiments the bias voltage cannot be perfectly fixed, however. It has fluctuations characterized by a distribution function $P(V)$ and what is actually measured is the average value

$$\langle \frac{dI}{dV} \rangle (\bar{V}) = \int_{-\infty}^{\infty} \frac{dI(V)}{dV} P(V) dV \qquad (2)$$

[8] where $\bar{V} = \int V P(V) dV$ is the average voltage. By inserting (1) into (2) and permuting integration orders, Equ. (2) can rewritten under the same form as (1) with the convolution product $P * f'_L$ in place of $f'_L$. Thus, not surprisingly, noise will further blur the resolution and appear under most circumstances as an excessive temperature. If $P(V)$ is Gaussian with an rms width $V_n$, the effective temperature can be evaluated as $T_{eff} = \sqrt{T^2 + 3(eV_n/k\pi)^2}$ [9].

**Experimental considerations**

Following the above discussion, in order to reach the highest level of spectroscopic resolution in a tunneling spectroscopy experiment it appears necessary to perform the experiment at the lowest possible temperature, which in practice means in a dilution refrigerator. It also appears equally important that the electromagnetic noise reaching the tunnel contact through the connecting wires has both a distribution and a spectrum compatible with the energy resolution sought. To reach such a low noise level in an experiment one must act at three levels. At the first level, the wires reaching the tunnel contact must be properly thermalized, so that the contact really "sees" an electromagnetic temperature equal to that of the refrigerator. This is achieved by careful shielding and filtering of the cables connected to the tunnel contact so that thermal radiation coming from room temperature apparatus is removed [10]. At the second level, the wiring of the experiment must follow well established low-noise techniques to avoid stray pick-up of electromagnetic signals in the laboratory ambient (mains, radio signals...) by the wires connected to the tunnel contact. In particular, if not using a magnetic shield enclosing the whole experimental setup, one has to select a proper circuit topology and grounding strategy to avoid noise due to inductive coupling in so-called ground loops [11]. At the third level, the intrinsic voltage noise of any instrument connected to the tunnel contact must be considered. Contrarily to the output noise of voltage generators which can usually be reduced to appropriate levels by using simple voltage dividers, the input noise of amplifiers is generally fully present on the tunnel contact in the bandwidth of the setup and can be seen as a back-action on the tunnel contact. Thus, in order to reach the ultimate energy resolution, one must use an amplifier with rms input voltage noise $V_n$ such that $V_n \ll kT/e$. As experiments are performed at lower temperatures, this puts more stringent requirements on the amplifier noise characteristics. Yet, in the low-temperature STM literature, this back-action is seldom considered and the current sensitivity (*i.e.* output voltage noise density divided by the transimpedance) is often the only figure of merit given to characterize the measurement chain.

**Implementation**

In our installation of an STM in a dilution refrigerator we have followed systematically the above considerations. The electromagnetic thermalization of all the wiring connected to the STM is achieved using shielded lines and custom microfabricated low-pass filters described in Ref. [10].

Besides this, we realized that having a very good control of the effective bias voltage applied to the tunnel contact is almost impossible when one uses a standard current-voltage converter (CVC) because such a circuit does not follow the good practice of having a single voltage reference point in a low signal level circuit. Namely, the voltage reference ("ground") of the CVC usually sitting at room temperature is physically separated from the electrode of the tunnel contact which should be precisely at the same potential. This makes the system prone to ground-loop noise problems. To overcome this problem we designed a modified CVC which allows differential sensing of the voltage on the tunnel contact. A simplified schematics of this amplifier in shown in Fig. 1.

In this design, the tip and sample "cold ground" are connected to the amplifier through a shielded twisted pair. The "cold ground" signal is sensed by an LT1028 ultra-low voltage noise op-amp and added to the desired tip bias voltage. This sum is then used as the voltage reference of the CVC. This design allows to cancel tip voltage fluctuations due to magnetic noise pick-up from the lab environment at low frequency which cannot be easily shielded. The current-voltage converter is made with an OPA627 which offers a good combination of current and voltage noises and bandwidth. Given the ~200 pF total capacitance to ground of the tunnel current line in the twisted pair cable, the 14 MΩ feed-back resistor and a ~1pF capacitor which limits output noise peaking at cut-off frequency [12], the OPA627 delivers a 22 kHz bandwidth, sufficient for STM imaging. The noise coming from the LT1028 is also limited by reducing its bandwidth using a 100 nF over-compensation capacitor and RC low-pass filter (neither shown). The operational amplifiers used in the rest of the circuit are low noise op-amps (OP27 or similar) which add negligible noise to the signal.

The amplifier is constructed using surface mounted components and fits easily on a 45 x 50 mm printed circuit board. The circuit board is mounted in a small metallic box which plugs directly into the socket of the twisted pair cable at the output of the refrigerator to avoid adding further cable length and stray capacitance. The amplifier is battery-powered and ground loops with other instruments are strictly avoided by using differential instrumentation amplifiers both in front of the bias input and following the output of the amplifier. The batteries and the instrumentation amplifiers are placed in an EMI shielding cabinet and a continuous metallic shield between the refrigerator and the cabinet completely encloses the amplifier and its connections to the batteries and the instrumentation amplifiers.

**Amplifier performance**
We used SPICE [13] simulations extensively during the design stage to optimize the level of noise of our amplifier and its dynamic behavior. SPICE simulations offer a rapid and simple way to evaluate several design options. This is particularly helpful for selecting the best suited operational amplifiers, as most manufacturers provide SPICE models for their op-amps and many of these models include noise characteristics. In Fig. 2 we show the predicted frequency dependences for key properties of our amplifier. The ac response to changes in the tunnel contact resistance and in the bias voltage are flat up to a -3 dB bandwidth of 22 kHz and 30 kHz respectively. The input impedance of the amplifier as seen from the tunnel contact is 14 Ω at very low frequency and then rises linearly with frequency up to a maximum of 10 kΩ at the cut-off frequency of the amplifier. The CVC thus delivers a good voltage bias at all frequencies for a tunnel contact in standard tunneling conditions (tunnel resistance $\gg h/e^2 \approx 26$ kΩ ). We have also investigated the efficiency of the differential voltage sensing scheme implemented in the amplifier at different frequencies. For this we simulated an ac voltage difference between the cold ground and the room-temperature ground and looked how much of this voltage was present on the tunnel contact bias (Fig. 2D). This shows that this setup can efficiently suppress low frequency noises due to currents flowing in the ground, and thus, provided the twisted pair wiring avoids picking up flux, it should be effective against pick-up of mains power and its harmonics.

In order to make accurate noise calculations, we need to properly take into account the noise due to the dissipative wires and filters sitting at different temperatures in our setup. Even though SPICE normally has a single temperature for all elements in the model, it is fairly easy to make a model for resistors at different temperatures for this purpose [14]. What the noise calculations show is that the input voltage noise of the amplifier, *i.e.* the back-action of the amplifier on the tunnel contact, is essentially determined by the input *voltage* noise of the CVC (OPA627) and resistor noise in the twisted pair and its filters [15]. The input voltage noise of the amplifier does not depend on the input *current* noise of the CVC, nor on the Johnson noise in the feed-back resistor which both only appear in the output noise of the OPA627 owing to the low input impedance of the CVC provided by feed-back loop. It neither depends on the LT1028 voltage sensor because its noise at the non inverting input of the OPA627 is negligible compared to the intrinsic voltage noise of the OPA627. For this setup, SPICE simulations predict that when running the experiment at 30 mK, the amplifier and wiring make a $\sim 2.0$ μV rms back action noise on the tip bias voltage. The smearing of the spectroscopic resolution due to this voltage noise has the same rms width than a 13 mK Fermi function, and its broadening of a 30 mK fermi function is similar to a 2.6 mK increase in temperature. Since this noise level is essentially due to the input voltage noise of the OPA627, it could be reduced by narrowing the bandwidth or by substituting it by an op-amp with less voltage noise. For this purpose one could use for instance an AD745, but its greater current noise would adversely affect the output noise of the CVC, *i.e.* its current sensitivity. As stated above, the output noise also depends on the Johnson noise of the feed-back resistor of the CVC. In fact, for the large resistor value we use, at room temperature this is the dominant output noise contribution at low frequency. Consequently, the sensitivity of the amplifier can be significantly improved by reducing the temperature of this resistor [16], as shown in Fig. 3. Using one extra coaxial line in the refrigerator, we placed the feed-back resistor at the temperature of the microscope (see Fig.1) but it could also be placed at 4 K with no change in performance. This, together with the twisted pair cable, makes a total of three lines for the tunnel contact bias and measurement. In this configuration, the output noise of the amplifier is peaked around the cutoff frequency of the amplifier, typical of an optimized CVC [12]. At low frequency, it rises due to $1/f$ noise. In between, around 100 Hz, the output noise has a flat minimum corresponding to a current sensitivity of 4 fA/Hz$^{1/2}$. This minimum is the most favorable frequency range for lock-in measurements of differential conductance since it provides the highest signal-to-noise ratio for the weak ac current one needs to measure when using an ac bias voltage well below $kT/e$ which is necessary to preserve the spectral resolution.

**Experimental result**

In Figure 4, we show a differential tunnel conductance measured with this amplifier. In this experiment the STM was equipped with a tungsten tip and the sample was a 25 nm-thick evaporated aluminium film deposited on an oxidized silicon substrate. The data were recorded at a temperature of $\sim 40$ mK, well below the superconducting transition temperature of the aluminium film, but above that of tungsten (15 mK). In this case, theory predicts that the differential conductance should be simply proportional to a convolution product of the BCS density of states of the film with the derivative of the fermi function of the normal state tip. The only adjustable parameters in this curve are the superconducting gap of aluminium [17], the asymptotic value of the conductance at large voltage and the temperature of the tip. Fitting the data with theory should thus give access to the effective temperature of the tip, including all noise contributions. In our case, for a reason not presently understood, a precise fitting of the data within this model is not possible. Nevertheless, the peak height in the differential conductance of such normal metal-superconductor contact gives a good indication of the effective temperature of the normal metal. Based only on the height of the peak, the effective temperature would be around 45 mK for a measurement performed at 40 mK. The fact that the peak height still varies when temperature and ac excitation are lowered confirms such an estimate and it also proves that the amplifier input noise is close to the predicted value. Thus, the spectroscopic resolution of this measurement is very close to the ultimate thermal limit and a factor of ~ 4 better than the best previously published resolution obtained with an STM.

**Acknowledgements**

We gratefully acknowledge discussions with G. Rubio-Bollinger and H. Courtois and useful comments of C. Urbina. This work was supported in part by the French Research Ministry project AC Nano 2003 NR110.

**Figures**

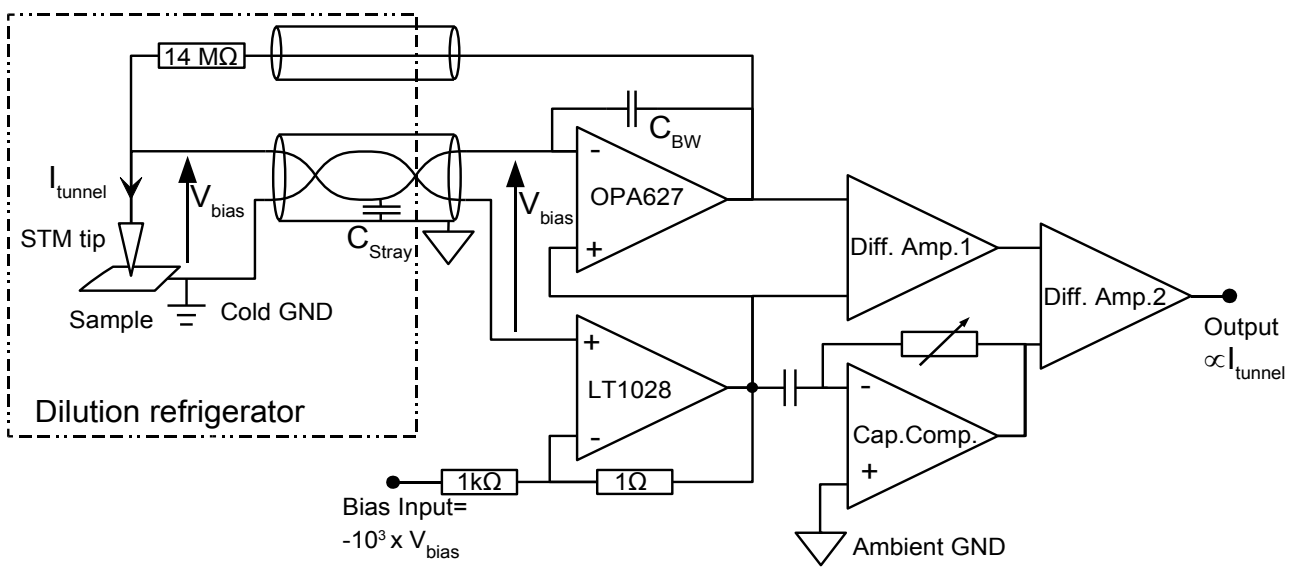

Fig. 1 Simplified schematics of the amplifier and connection to the tunnel contact. The tip and sample « cold ground » are connected to the amplifier through a shielded twisted pair with filters (not shown). The « cold ground » signal is sensed by an LT1028 ultra low voltage noise op-amp and used to offset the desired tip bias voltage. This sum is then used as the voltage reference of the current voltage converter. The first difference amplifier (Diff. Amp. 1) is made using high precision resistors (0.01%) to accurately subtract the bias voltage from the output of the CVC. The two other circuits (Diff. Amp. 2 and Cap. Comp.) compensate the ac currents flowing to ground in the stray capacitance $C_{Stray}$~200 pF. For best current sensitivity, the 14 MΩ feed-back resistor of the CVC is placed at low temperature and connected through a coaxial cable. Bandwidth of the CVC is adjusted to 22 kHz to limit noise peaking by the use of $C_{BW}$~1pF.

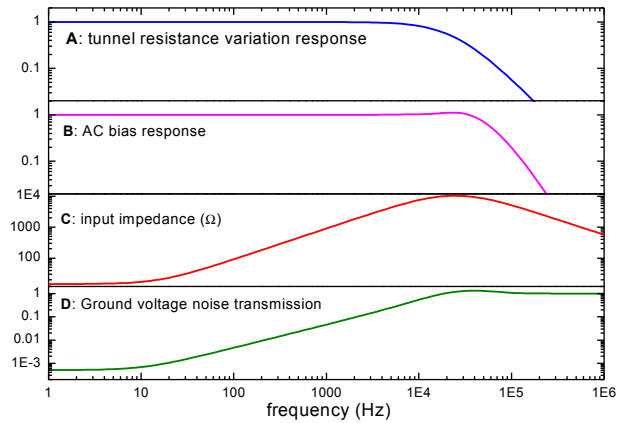

Fig. 2
Results of SPICE simulations. Panel A: normalized linear ac response at the output of the amplifier in response to a small variation of the tunnel resistance, as a function of frequency. The normalization value is 1.4 µV per % variation of a 1 MΩ tunnel resistance under a dc bias of 10 µV. Panel B: normalized linear ac response at the output of the amplifier in response to ac bias. The normalization value is 14 µV per µV ac bias on a 1 MΩ tunnel resistance. Panel C: Input impedance of the CVC as seen from the tunnel contact. Panel D: Assuming an ac voltage difference between hot and cold grounds, fraction of which still present in the tunnel contact bias.

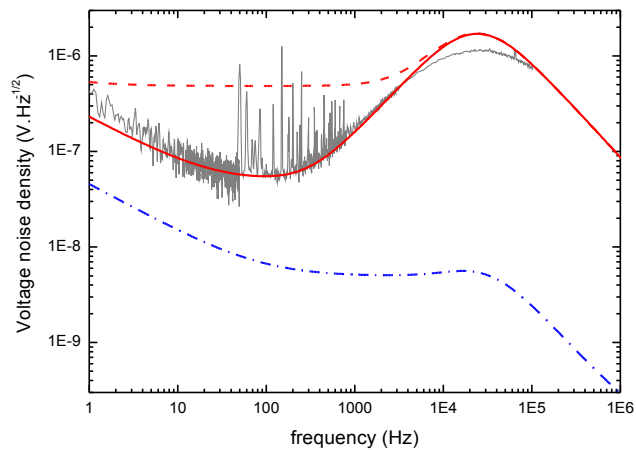

Fig. 3
Noise performance of the amplifier. Dash-dotted curve : Predicted input noise voltage density on the tunnel contact. The input noise is independent of the feed-back resistor temperature in this range. The integrated noise gives ~2 µV rms. Thick top curves : predicted output noise with the feed-back resistor at room temperature (dashed line) or at low temperature (solid line). Thin gray line: measured output noise while experiment was at low temperature. No adjustable parameters are used. The discrepancy above 10 kHz is attributed to an additional stray feedback capacitor on the OPA627 in parallel with the $C_{BW}$=1 pF installed capacitor (see Fig. 1).

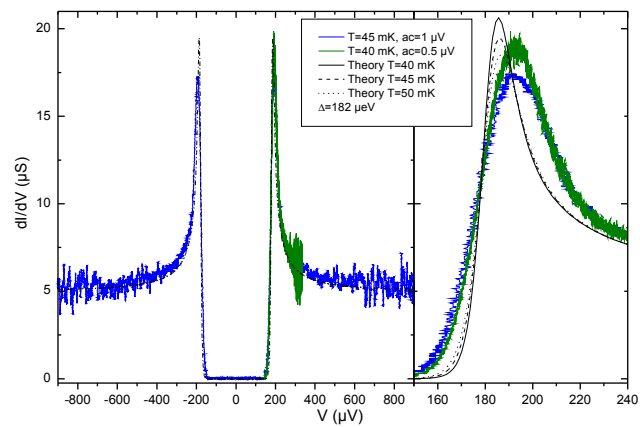

Figure 4.
Experimental differential conductance taken on a 25 nm thick aluminium film with a tungsten tip for two different temperatures and ac excitations, and comparison with the prediction for Normal-Superconductor tunnel spectrum at different temperatures, with a gap Δ=182 μeV. On the right panel, a close-up on the right hand side peak. In spite of the unexplained discrepancy, the aspect ratio of the peak and its dependence on ac amplitude and/or temperature in this range indicate that the effective temperature of the measurement is close to the refrigerator temperature and proves that the back-action of our amplifier is not larger than expected from the simulations.